# How AI Agents Follow the Herd of AI? Network Effects, History, and Machine Optimism

**Yu Liu**
Fudan University
yuliu23@m.fudan.edu.cn

**Wenwen Li**
Fudan University
liwwen@fudan.edu.cn

**Yifan Dou**
Fudan University
yfdou@fudan.edu.cn

**Guangnan Ye**
Fudan University
yegn@fudan.edu.cn

**Abstract:** *Understanding decision-making in multi-AI-agent frameworks is crucial for analyzing strategic interactions in network-effect-driven contexts. This study investigates how AI agents navigate network-effect games, where individual payoffs depend on peer participation—a context underexplored in multi-agent systems despite its real-world prevalence. We introduce a novel workflow design using large language model (LLM)-based agents in repeated decision-making scenarios, systematically manipulating price trajectories (fixed, ascending, descending, random) and network-effect strength. Our key findings include: First, without historical data, agents fail to infer equilibrium. Second, ordered historical sequences (e.g., escalating prices) enable partial convergence under weak network effects but strong effects trigger persistent "AI optimism"—agents overestimate participation despite contradictory evidence. Third, randomized history disrupts convergence entirely, demonstrating that temporal coherence in data shapes LLMs' reasoning, unlike humans. These results highlight a paradigm shift: in AI-mediated systems, equilibrium outcomes depend not just on incentives, but on how history is curated, which is impossible for human.*

## 1. Introduction

The study of strategic decision-making through game-theoretic frameworks has long been a cornerstone of understanding agent behavior in interactive environments. While classic games such as the Prisoner's Dilemma and negotiation games have been replicated and analyzed in multi-agent systems (Fan et al., 2024), far less attention has been paid to scenarios where individual payoffs are intrinsically tied to network effects—the phenomenon where an agent's utility depends on the number of peers adopting the same strategy. Such scenarios mirror real-world coordination challenges, from technology adoption to social participation, where value is dynamically shaped by collective behavior. Unlike traditional games with static equilibria, these settings demand agents to engage in recursive reasoning about others' beliefs and actions, creating layers of strategic complexity. This raises a critical question: How do LLM agents navigate such interdependencies when they cannot practically compute levels of recursion of others, and how do their assumptions about others' computational capabilities shape collective outcomes?

Our research first examines a repeated network-effect game where AI agents decide whether to "participate," with payoffs affected by peer participation. Then, we propose a novel workflow design inspired by theoretical findings from network effect literature to incorporate historical data—encoded as price-participation trajectories—into agents, aiming to assist agents' decision making. We show that the organization of history critically shapes network dynamics. Crucially, as network effects intensify, agents increasingly diverge from theoretical equilibrium documented in classical economic models. Unlike humans, who experience history as an immutable linear sequence, LLM agents learn from history as malleable data—filtered, reordered, or artificially curated—which fundamentally reshapes their strategic expectations. Through experiments that alter how historical trajectories are formatted and injected (e.g., emphasizing selective interactions), we demonstrate that LLM agents' expectations about peers depend not just on what happened, but how the past is computationally framed. This plasticity establishes a new frontier in understanding game theory of AI agents: the design of history itself emerges as a strategic variable, with profound implications for AI systems in socially embedded, history-sensitive environments.

## 2. A canonical example of network-effect games

A canonical example of network-effect games involves six scholars deciding whether to attend a conference. Each scholar has full knowledge of the game structure, including the set of players, the action set, and the generic payoff function. Specifically, each scholar $j$ knows:

- The total number of scholars ($n = 6$)
- The action set for all scholars: $\{Attend, Not\ Attend\}$.
- Individual parameters: standalone value $\theta_j$ (ranging from 1 to 6 across agents), a coefficient $\beta$ that measures the strength of network effects, and a fixed cost $p_j$ that measures traveling expenses (e.g., airfare, registration).
- $U_j$: The payoff function for each scholar $j$, where each scholar will choose to attend the conference if her/his utility is non-negative. $U_j$ is defined as:

$$U_j(\theta_j) = \theta_j + \beta N - p_j \geq 0, \tag{1}$$

in which, $N$ is the total attendees that is unobservable ex ante, requiring agents to form expectations about others' choices (Katz and Shapiro 1985). Economic theory resolves this circularity with the *fulfilled expectation equilibrium (FEE)*: agents coordinate on a shared expectation of $N$, which aligns exactly with the realized attendance. While elegant, FEE assumes perfect homogeneity in human reasoning. By replicating this scenario with LLM-based agents—whose access to historical data can be algorithmically manipulated—we test whether machine-driven systems adhere to or deviate from FEE predictions. This also offer insights into other network-effect-related contexts like transportation systems or financial markets, where the presence of AI agents is increasing.

## 3. Experiment Setting

To simulate the 6-scholar game using LLM agents, we utilize three state-of-the-art LLMs from Alibaba Cloud's Qwen family as the backbone: Qwen-max (high-performance model with advanced reasoning), Qwen-turbo (general-purpose medium model), and Qwen-2.5-1.5B (lightweight benchmark). This hierarchy allows us to test how parameter size and specialized training—critical for mathematical and strategic reasoning—affect coordination. For all LLMs, we set the temperature to 0.7. Prior to experiments, we validated each model's comprehension of the conference-attendance rules through iterative prompting, ensuring alignment with the game's utility structure and equilibrium logic.

Our experiment features two roles: a manager (e.g., "conference chair") that controls game flow, sets prices, $p$, records agent decisions, and computes participation outcomes, and six LLM-based scholar agents tasked with expecting peer attendance. Each agent knows their standalone value $\theta$ (1–6) and applies Equation (1) to decide participation but cannot observe peers' real-time choices. Critically, between rounds, agents receive only two updates from the manager: the new price $p$ and the aggregate historical participation count from prior rounds. This design mirrors real-world scenarios where actors gauge collective action without full visibility into individual decisions.

The experiment unfolds in two phases: setup and play. In setup, network-effect strength $\beta$, standalone values $\theta$ (assigned as 1–6 across agents), and initial price $p$ are configured. Each round proceeds iteratively: 1) The manager sets a new $p$ for the round; 2) Agents independently make the expectation on peer participation, $N$, without observing peers' real-time choices; 3) The manager aggregates participation decisions, computes payoffs, and shares only the total historical participation count with agents before the next round.

## 4. Workflow design on historical information

Inspired by theoretical insights from network effect literature, we propose a novel workflow design to systematically incorporate historical information into agents, aiming to enhance agents' reasoning abilities and their rational behavior. Our workflow design embeds historical context into agents' decision processes, allowing iterative learning from past participation and outcomes. Notably, this design aligns conceptually with the in-context learning abilities observed in LLMs, where LLMs can learn from prior knowledge or task-relevant examples.

To investigate how historical information shapes decision-making, we test two configurations: 1) Non-Repeated Game (Static): Agents face a one-shot game with no historical data, iterated 10 times per price. This isolates agents' initial reasoning in a static environment. 2) Repeated Game (Dynamic): Games run recursively with evolving prices. Agents observe their own past expectations, participation outcomes, and derived payoffs from learning from prior rounds. The static game isolates "first-reaction" behavior, while the dynamic setting tests how agents refine expectations iteratively using history—a critical distinction since LLMs, unlike humans, can algorithmically reinterpret past data.

Central to our workflow design in repeated games are price trajectories, derived from theoretical participation levels (1–6 agents) to test four pricing strategies: *fixed*, *ascending*, *descending*, and *randomized prices*. Each trajectory manipulates how agents encounter costs over rounds, with the manager adjusting prices six times per round (one per theoretical participation threshold) and collecting agent expectations at each step. This design probes how price sequencing—stable, escalating, de-escalating, or chaotic—shapes iterative belief formation. To ensure consistency with the non-repeated baseline, all trajectories are independently repeated 10 times, isolating the interplay between history's structure (price paths) and agents' adaptive reasoning under network effects.

## 5. Results and Discussion

This section presents experiment results from benchmark tests (Qwen-2.5-1.5B) and workflow evaluations (Qwen-turbo) across four pricing strategies. Additional experiments employing Qwen-max yielded consistent results with the Qwen-turbo trials. Due to space constraints, detailed results from Qwen-max experiments will be available upon request.

**5.1 Benchmarks (Static Game)**

Figure 1 establishes benchmarks for the static, non-repeated game across two network-effect regimes: weak ($\beta$=0.25, Panel a) and strong ($\beta$=0.75, Panel b). In both panels, red lines depict the theoretical FEE, where participation declines as prices rise—a pattern intuitive to human reasoning (higher cost, fewer participants). However, AI agents exhibit stark deviations: without historical data, their expectations (box plots) show significant dispersion, failing to converge toward the theoretical trend. Notably, even as prices increase, mean agent expectations remain disconnected from the FEE curve, revealing an inability to internalize the causal link between price and participation. This divergence persists regardless of $\beta$, demonstrating that LLM agents—even the state-of-the-art models—lack the inherent capacity to simulate recursive reasoning in network-effect contexts.

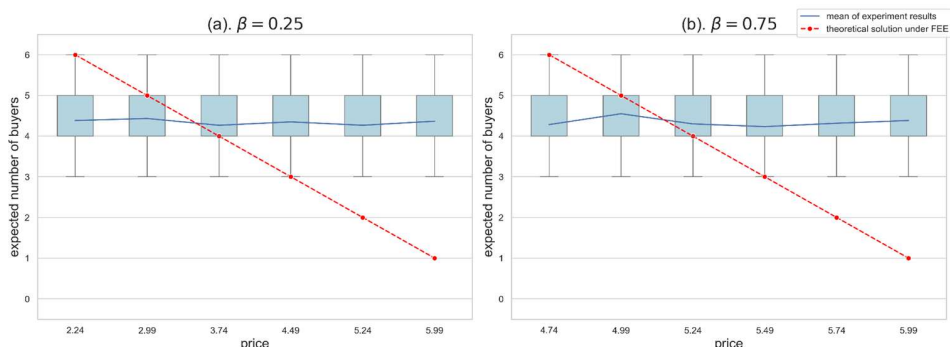

Figure 1. Benchmarks (red lines represent the theoretical value)

**5.2 Fixed Price**

Figure 2 examines the repeated-game setting under fixed prices, again contrasting weak (Panels a–c) and strong (Panels d–f) network effects. In Panels 2(a)–2(c), agents gradually converge toward theoretical participation levels ($N$=6,4,2) as they observe historical outcomes. For example, in Panel 2(a) (price $p$=2.24), agents align with $N$=6 within three rounds, while in Panel 2(b) ($N$=4), convergence is similar. However, in Panel 2(c) ($N$=2), persistent dispersion in expectations (evidenced by box plots) reveals residual optimism: agents struggle to accept low participation equilibria even with unambiguous historical evidence. This suggests LLMs inherently favor coordination optimism.

This "AI optimism" becomes destabilizing under stronger network effects (Panels d–f). Despite receiving historical data, agents fail to converge to theoretical equilibrium ($N$=6,4,2) when $\beta$=0.75. For instance, in Panel 2(d), expectations remain inflated above $N$=6, even though historical payoffs should signal over-participation. The amplified network benefits ($\beta$) appear to override rational inference from past outcomes, as agents prioritize the potential for collective gains over evidence of individual losses. This divergence underscores a critical limitation: LLMs' reasoning in networked environments is disproportionately blurred by the magnitude of interdependencies.

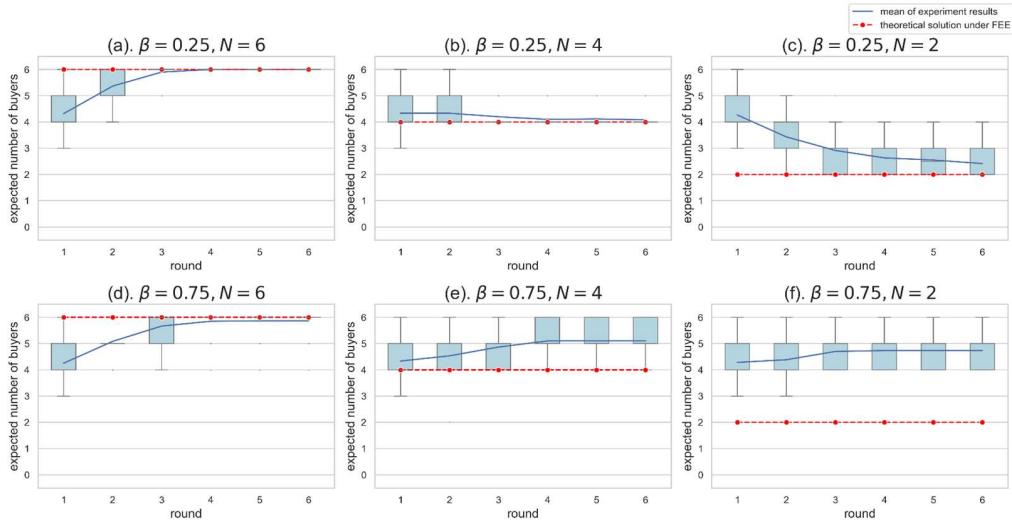

Figure 2. Convergence to Theoretical Value under The Dynamic Setting with The Same Price

**5.3 Increasing Prices**

Figure 3 analyzes agents' expectations under escalating price trajectories, comparing weak ($\beta=0.25$, Panel a) and strong ($\beta=0.75$, Panel b) network effects. In Panel a, prices rise from 2.24 to 5.99, mirroring the theoretical FEE's inverse price-participation relationship (red line). Initially, agents exhibit the same behavior from Figure 1, displaying dispersed and random expectations at the lowest price. However, as rounds progress and historical participation data accumulates, their mean expectations (blue line) slope downward, aligning with the theoretical trend. This indicates that history enables LLMs to infer price effects, despite their initial reasoning deficits.

However, this learning breaks down under strong network effects (Panel b). Here, agents persistently overestimate participation (blue line deviates upward from red FEE), even as prices surpass thresholds that should deter attendance. Again, the high network benefit fuels "AI optimism"—agents prioritize the potential gains from peer participation over historical evidence of declining payoffs. For example, at p=5.99, where FEE predicts only 1 participation, agents still expect substantial turnout (N≈4), defying both logic and historical feedback. This suggests that in strongly networked systems, LLMs' reasoning is overridden by the perceived value of collective action, rendering price signals ineffective.

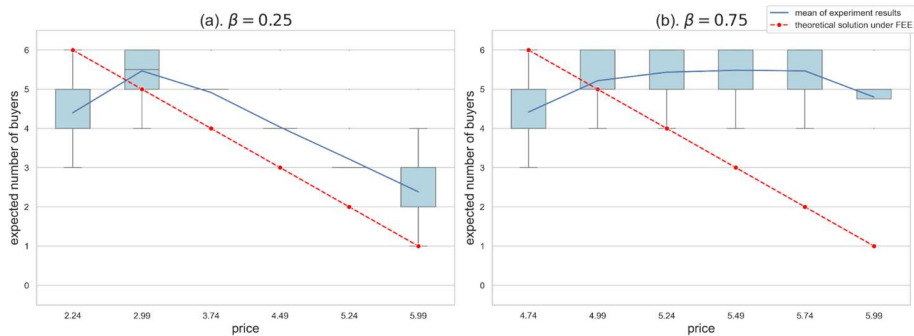

Figure 3. Increasing Price-Participation under Network Effects

### 5.4 Decreasing Prices

Figure 4 analyzes descending price trajectories, starting with high costs ($p$=5.99) and decreasing over rounds. Initially, agents exhibit behavior akin to Figure 1 at high prices. However, as prices decline, convergence improves markedly. In Panel a, agents align with FEE when prices are substantially low. Strikingly, in Panel b's left-hand side (lowest prices), even with strong network effects ($β$=0.75), expectations converge toward higher participation levels. This reflects a confluence of forces: smaller prices reduce individual costs, while stronger network effects amplify perceived collective benefits. The result is a "sweet spot" where structural incentives (low $p$, high $β$) align with AI optimism, masking whether convergence stems from rational learning or mere coincidence of favorable conditions. This duality complicates interpretations of LLM reasoning, suggesting that apparent alignment with theory may not always signal true strategic understanding.

### 5.5 Random Price

Lastly, Figure 5 tests randomized price trajectories, revealing that the ordering of history—not just its existence—critically shapes LLM agents' reasoning. Under both weak (Panel a) and strong (Panel b) network effects, agents fail to converge toward theoretical equilibrium when prices lack a coherent sequence. Unlike structured trajectories (ascending/descending), random price fluctuations obscure causal relationships between cost and participation, leaving agents unable to generalize patterns from disjointed historical snapshots. This result highlights that LLMs' capacity to "learn" depends heavily on temporal coherence in data—a constraint absent in human reasoning, where causal inference persists despite noise.

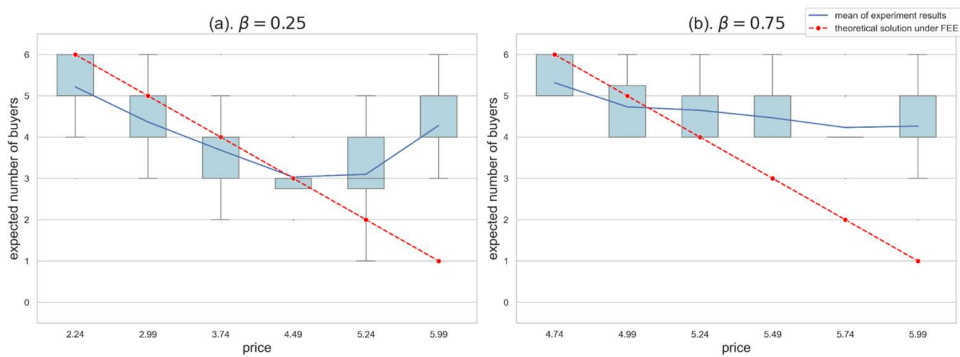

Figure 4. Decreasing Price-Participation under Network Effects

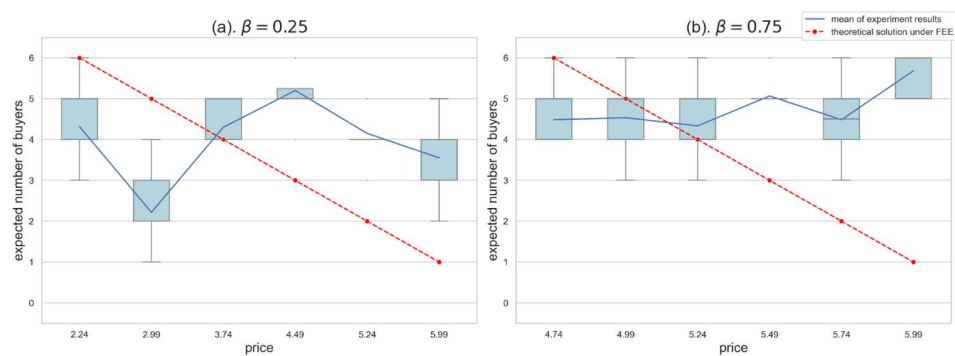

Figure 5. Randomly-Assigned5 Price-Participation under Network Effects

## 6. Conclusion

Our experiments reveal that LLM agents' ability to navigate network-effect games hinges not just on access to historical data, but on how that history is structured. While ordered price trajectories (ascending/descending) enable partial alignment with theoretical equilibria under weak network effects, strong interdependencies foster persistent "AI optimism" and override rational inference. These findings challenge classical equilibrium models in the context of AI-mediated systems, which implies the need to design history-aware frameworks that account for machines' unique "cognitive" biases. This work represents only the beginning of this line of research, as our focus on a controlled setting only contains fixed agent roles. Future work should explore hybrid human-AI networks and real-world data noise.